\documentstyle[12pt]{article}
\newlength{\extraspace}
\newlength{\extraspaces}
\newcommand{\be}{\begin{equation}
\addtolength{\abovedisplayskip}{\extraspaces}
\addtolength{\belowdisplayskip}{\extraspaces}
\addtolength{\abovedisplayshortskip}{\extraspace}
\addtolength{\belowdisplayshortskip}{\extraspace}}
\newcommand{\ee}{\end{equation}}
\newcommand{\ba}{\begin{eqnarray}
\addtolength{\abovedisplayskip}{\extraspaces}
\addtolength{\belowdisplayskip}{\extraspaces}
\addtolength{\abovedisplayshortskip}{\extraspace}
\addtolength{\belowdisplayshortskip}{\extraspace}}
\newcommand{\ea}{\end{eqnarray}}
\newcommand{\nonu}{\nonumber \\[.5mm]}

\newcommand{\e}{\, {\rm e}}
\newcommand{\D}{{\cal D}}

\newcommand{\ket}[1]{\left\vert {#1} \right\rangle}
\newcommand{\VEV}[1]{\left\langle {#1} \right\rangle}

\newcommand{\A}{&\!\!\!}

\begin{document}
\addtolength{\baselineskip}{.7mm}
\thispagestyle{empty}
\begin{flushright}
TIT/HEP--204 \\
STUPP--92--131 \\
August, 1992
\end{flushright}
\vspace{2mm}
\begin{center}
{\large{\bf Correlation Functions in \\[2mm]
Two-Dimensional Dilaton Gravity}} \\[9mm]
{\sc Yoichiro Matsumura, Norisuke Sakai} \\[3mm]
{\it Department of Physics, Tokyo Institute of Technology \\[2mm]
Oh-okayama, Meguro, Tokyo 152, Japan} \\[4mm]
{\sc Yoshiaki Tanii} \\[3mm]
{\it Physics Department, Saitama University \\[2mm]
Urawa, Saitama 338, Japan} \\[4mm]
and \\[4mm]
{\sc Taku Uchino} \\[3mm]
{\it Department of Physics, Tokyo Institute of Technology \\[2mm]
Oh-okayama, Meguro, Tokyo 152, Japan} \\[9mm]
{\bf Abstract}\\[5mm]
{\parbox{13cm}{\hspace{5mm}
The Liouville approach is applied to the quantum treatment of the
dilaton gravity in two dimensions. The physical states are obtained
from the BRST cohomology and correlation functions are computed up
to three-point functions. For the $N=0$ case (i.e., without
matter), the cosmological term operator is found to have the
discrete momentum that plays a special role in the $c=1$ Liouville
gravity. The correlation functions for arbitrary numbers of
operators are found in the $N=0$ case, and are nonvanishing only
for specific ``chirality'' configurations.}}
\end{center}
\vfill
\newpage
\setcounter{section}{0}
\setcounter{equation}{0}
%\addtolength{\baselineskip}{2mm}
%
%%%%%%%  Introduction  %%%%%%%%%%%%%%%%%%%%%%%%%%%%%%%%%%%%%%%%%%%
%
The two-dimensional gravity interacting with a dilaton field and
matter fields has attracted much attention recently ever
since the work on the black hole evaporation \cite{CGHS}.
Many efforts are devoted to study the Hawking radiation and the
black hole evaporation in the model \cite{CGHS}--\cite{BICA}.
Most of the works have eventually employed the semi-classical
approximation, which is often blamed to be the
possible origin of diseases in this problem.
Therefore it is very desirable to have a full quantum treatment of
the dilaton gravity even for a restricted class of models.
In two dimensions, we can perhaps hope to understand
quantum theory of the dilaton gravity without using the
semi-classical approximations.
The continuum approach of the Liouville theory is most suitable in
discussing the dilaton gravity, since the theory is nonlinear and
it is difficult to invent the discretized version of the model
such as matrix models.
\par
The purpose of our paper is to determine the physical states and
the gauge invariant operators by a BRST analysis and compute
their correlation functions in the dilaton gravity coupled to
$N$ massless free scalar fields applying the methods used in the
Liouville theory.
We shall take the recently proposed models of dilaton
gravity \cite{BICA} that are conformally invariant.
\par
We obtain the BRST cohomology for the case $N\ge 24$ that
we cannot find in the existing literature.
Apart from the usual string states with momentum and oscillator
excitations, we find that there are only a few physical states
with nontrivial ghost numbers. This situation is very similar to
the case of $0<N<24$ analyzed in ref.\ \cite{BILAL}, but is in
sharp contrast to the case of $N=0$ which is essentially the same
system as the $c=1$ two-dimensional gravity, except that one of
the field is of negative metric. Similarly to ref.\ \cite{BICA},
we determine the cosmological term by demanding that it should be
a gauge (BRST) invariant operator which reduces to the classical
cosmological term in the weak coupling limit. The cosmological term
in the case of $N=0$ is particularly interesting. We find that the
momentum of the cosmological term is precisely at the smallest of
the discrete momenta where the characteristic symmetry structure
of the $c=1$ two-dimensional gravity has been observed as the
ground ring \cite{WITTEN}.
\par
By applying the methods used in the Liouville theory \cite{GOULI},
\cite{DFKU} we compute the correlation functions of gauge (BRST)
invariant operators with momentum.
The correlation functions up to three gauge invariant operators can be
obtained for the general $N>0$ case, and exhibit singularity
structures expected from the physical state spectrum.
For the $N=0$ case (without matter fields), we obtain correlation
functions for arbitrary
numbers of gauge invariant operators and find that the correlation
functions have a singular factor which can be absorbed as a
renormalization of the cosmological constant. Since the
cosmological term has nonvanishing momentum with positive
``chirality,'' we obtain an interesting selection rule with respect to
chirality. The correlation functions are nonvanishing in the case
where only one of the gauge invariant operators has negative
chirality and all the other operators have positive chirality.
All other chirality configurations give vanishing correlation
functions.
\par
Let us explain the relevance of the correlation functions to the
quantum treatment of the dilaton gravity on the two-dimensional
spacetime, such as the black hole evaporation.
An insertion of these gauge invariant operators
corresponds to creating a hole with appropriate boundary conditions
in the language of the physics on the two-dimensional spacetime.
If our analysis is extended to macroscopic
loops from the local operators, we can find quantum transitions
leading to topology change in the two-dimensional dilaton gravity.
In order to discuss the black hole evaporation, we need to prepare
the appropriate wave function for the black hole and to examine its
evolution. We shall use the momentum eigenstates as the gauge
invariant operators to compute correlation functions. From the
two-dimensional physics point of view, the momentum is just a
conserved charge associated with the one dimensional ``universe.''
\par
%
%%%%%%%  The conformally invariant model  %%%%%%%%%%%%%%%%%%%%%%%%%%%%%
%
First we consider a two-dimensional manifold with a metric
$\bar g_{\mu\nu}$ whose signature is Euclidean.
The Einstein action is a topological invariant in two dimensions.
However, it acquires a dynamical meaning when multiplied by a
function of a scalar field $\phi$ which we parametrize as
$\e^{-2\phi}$
\ba
S_{\rm classical} \A = \A S_{\rm dilaton} +S_{\rm matter}, \qquad
S_{\rm dilaton} = {1 \over 2\pi} \int d^2 z \sqrt{\bar g} \,
\left[ - \e^{-2\phi} \bar R + 2 \mu \right], \nonu
S_{\rm matter} \A = \A {1 \over 8\pi} \int d^2 z \sqrt{\bar g}
\sum_{j=1}^N \bar g^{\mu\nu} \partial_\mu f^j \partial_\nu f^j,
\label{noderdilaton}
\ea
where the scalar field $\phi$ is called dilaton and $N$ free
massless scalar fields $f^j$ are chosen as matter fields.
A function of scalar fields multiplying the
Einstein action can be absorbed into the metric by a local
Weyl transformation in spacetime dimensions other
than two. Therefore the dilaton has a special status in two
dimensions: it cannot be eliminated by
local Weyl transformations and the Einstein action without the
dilaton field is a topological invariant that is dynamically
empty. To clarify the significance of the dilaton field
more clearly, we can make a local Weyl transformation
$g_{\mu\nu} = \e^{2\phi} \bar g_{\mu\nu}$
\be
S_{\rm dilaton} = {1 \over 2\pi} \int d^2 z \sqrt{g} \e^{-2\phi}
\left[ - R -4 g^{\mu\nu} \partial_\mu \phi \partial_\nu \phi
+ 2 \mu \right].
\label{cghsaction}
\ee
This form of the dilaton gravity system is suggested by the
string theory and has been extensively discussed in connection
with the black hole evaporation \cite{CGHS}--\cite{BICA}.
In the same spirit, we shall also regard this metric
$g_{\mu\nu}$ as physical.
\par
Recently an approach using the conformal field theory was proposed
to discuss the quantization of the dilaton gravity
\cite{BICA}. We shall take this approach
and give a slight generalization of their models.
We can borrow the idea of string theories to quantize the dilaton
gravity theory in two dimensions.
In particular, the partition function $Z$ is given as a sum over
various topologies, and the weight is given by powers of the
topological expansion parameter $g_{\rm st}$ (string expansion
parameter), in accordance with the idea of string theories.
The contribution $Z_h$ from the Riemann surface with $h$ handles
is given by a path integral
\be
Z = \sum_{h=0}^{\infty} g_{\rm st}^{2h-2} Z_h, \qquad
Z_h = \int {\D_g g_{\mu\nu} \D_g \phi \D_g f \over
V_{\rm gauge}} \, \e^{- S_{\rm classical}[g, \phi, f]},
\label{partfunc}
\ee
where $V_{\rm gauge}$ is the volume of the group of
diffeomorphisms. We use the conformal gauge $g_{\mu\nu} =
\e^{2\rho}\hat g_{\mu\nu}$ with the Liouville field $\rho$ and
the reference metric $\hat g_{\mu\nu}(\tau)$ that depends on the
moduli parameters $\tau$ of the Riemann surface.
The functional integral over the metric
can be converted into an integral over the ghosts
$c^\mu,\ b_{\mu\nu}$, the Liouville field and the moduli,
which is divided by the volume $V_{\rm CKV}$ of the group
generated by conformal Killing vectors  \cite{POLYAKOV}.
\par
We have used the physical metric $g_{\mu\nu}$ to define the
functional measure in quantizing the matter and the ghost.
It is convenient to transform the measure into a measure using
$\hat g_{\mu\nu}$. The Jacobian is given by the Weyl anomaly which
is expressed by the Liouville action $S_{\rm L}$
\cite{POLYAKOV}
\ba
\D_{\e^{2\rho} \hat g} f \D_{\e^{2\rho} \hat g} b
\D_{\e^{2\rho} \hat g} c
\A = \A \D_{\hat g} f \D_{\hat g} b \D_{\hat g} c \,
\e^{- S_{\rm anomaly}^{{\rm gh}, f}}, \nonu
S^{{\rm gh}, f}_{\rm anomaly}
= {26-N \over 12} S_{\rm L}[\rho, \hat g]
\A = \A {26-N \over 24\pi} \int d^2 z \sqrt{\hat g}
\left( \hat g^{\mu\nu} \partial_\mu \rho \partial_\nu \rho
+ \hat R \rho \right).
\label{anomalymatter}
\ea
Since the measure for the Liouville field is nonlinear,
the Jacobian for the transformation to the translation invariant
measure using $\hat g_{\mu\nu}$ is more difficult to determine.
It has been argued that this Jacobian must be a local functional
of the Liouville field $\rho$ containing a bilinear kinetic term,
a term linear in the curvature and $\rho$, and the exponential
term representing the cosmological term \cite{DDK}.
This is an extremely successful Ansatz.
The quantization of the dilaton is similar but may be more complicated.
There has also been a proposal for the use of metric
${\rm e}^{-2\phi}g_{\mu \nu}$ rather than $g_{\mu \nu}$ in defining
the ghost measure \cite{STR}. Therefore, along the spirit of
ref.\ \cite{DDK}, we take the following more
general Ansatz for the Jacobian for the transformation of the
functional measure to the translationally invariant measure using
$\hat g_{\mu\nu}$ in the case of the dilaton gravity with
the matter
\ba
S \A = \A S_{\rm kin} + S_{\rm cosm}, \qquad
S_{\rm kin}  =  S_{\rm dilaton}(\mu=0) + S_{\rm anomaly}
+ S_{\rm matter}, \nonu
S_{\rm kin}
\A = \A {1 \over 2\pi} \int d^2 z \sqrt{\hat g}
\biggl[ {\rm e}^{-2\phi} \left(
-4 \hat g^{\mu\nu} \partial_\mu \phi \partial_\nu \phi
+ 4 \hat g^{\mu\nu} \partial_\mu \phi \partial_\nu \rho
- \hat R \right) \nonu
\A\A - \kappa \left( \hat g^{\mu\nu} \partial_\mu \rho
\partial_\nu \rho + \hat R \rho \right)
+ a \left(2 \hat g^{\mu\nu} \partial_\mu \phi \partial_\nu \rho
+ \hat R \phi \right) \nonu
\A\A + b \hat g^{\mu\nu} \partial_\mu \phi \partial_\nu \phi
+ {1 \over 4} \sum_{j=1}^N
\hat g^{\mu\nu} \partial_\mu f^j \partial_\nu f^j \biggr].
\label{kinetic}
\ea
This is the most general Ansatz for the anomaly under the
assumption that the measures are defined by
${\rm e}^{\alpha\phi}g_{\mu\nu}$ for the quantization of various
fields\footnote{If the amount of the anomaly $\gamma_j$ is defined
by ${\rm e}^{2\alpha_j\phi}g_{\mu\nu}$, the parameters are given by
$a=\sum\gamma_j\alpha_j$, $b=\sum\gamma_j\alpha_j^2$. If we allow
the coefficients of terms linear in $\hat R$ to be arbitrary
($-\kappa' \hat R\rho, \kappa'\not=\kappa$ and
$a'\hat R\phi, a'\not=a$),
we cannot make the action into a free field form by the change of
variables like (\ref{fieldredef}).}.
This is a slight generalization of the Ansatz in refs.\ \cite{STR},
\cite{BICA}. The cosmological term $S_{\rm cosm}$
will be determined after we discuss the physical states.
\par
We can reduce the above kinetic term to a free field action by a
change of variables.
In the case of $\kappa\not=0$, the change of variables reads
\ba
\omega \A = \A {\rm e}^{-\phi}, \quad
\chi = -{\rho \over 2}-{\omega^2 + a\ln\omega \over 2\kappa}
={\hat \chi \over 4\sqrt{|\kappa|}}, \nonu
d\Omega \A = \A -{1 \over \kappa}
d\omega \sqrt{\omega^2 - \kappa + a
+ {a^2 + b\kappa \over 4\omega^2}}
={d\hat \Omega \over 4\sqrt{|\kappa|}}.
\label{fieldredef}
\ea
One should note that the Liouville field $\rho$ in the original
metric is contained only in $\hat \chi$. Therefore
$\hat \Omega$ is just a redefined dilaton field and transforms as a
genuine scalar field under general coordinate transformations,
whereas the field $\hat \chi$ transforms as the
Liouville field (conformal factor of the metric).
This change of variables (\ref{fieldredef})
gives a free field action with the source term for $\hat \chi$
\be
S_{\rm kin} = {1 \over 8\pi} \int d^2 z \sqrt{\hat g} \biggl( \mp
\hat g^{\mu\nu} \partial_\mu \hat \chi \partial_\nu \hat \chi
\pm 2 \sqrt{|\kappa|} \hat R \hat \chi
\pm \hat g^{\mu\nu} \partial_\mu \hat \Omega
\partial_\nu \hat \Omega + \sum_{j=1}^N
\hat g^{\mu\nu} \partial_\mu f^j \partial_\nu f^j \biggr),
\label{dilkin}
\ee
where the upper (lower) signs are for $\kappa > 0$ ($\kappa < 0$).
The free field theory is almost identical to the usual Liouville
theory with matter fields, except that
there is one field with negative metric.
The transformed Liouville field $\hat \chi$ has negative metric
in the case of $\kappa>0$, whereas the transformed dilatonic
field $\hat \Omega$ has negative metric in the case of $\kappa<0$.
For $\kappa = 0$, the appropriate change of variables is
given by\footnote{The eq.(\ref{fieldredefzero}) has the
$a \rightarrow 0$ limit and is valid for $a=0$ too.}
\be
\chi^\pm = Q \left(- \rho - \frac{b}{2a} \phi
- \frac{2a+b}{4a} \ln \left| \e^{-2\phi} + \frac{a}{2} \right|
\right) \pm \frac{2}{Q} \left( \e^{-2\phi} - a\phi \right).
\label{fieldredefzero}
\ee
In this case, $\chi^+ + \chi^-$ transforms as the
Liouville field, while $\chi^+ - \chi^-$ transforms as a scalar
field. The free field action reads
\ba
S_{\rm kin} \A = \A {1 \over 8\pi} \int d^2 z \sqrt{\hat g} \,
\biggl( \hat g^{\mu\nu} \partial_\mu \chi^+ \partial_\nu \chi^+
- Q \hat R \chi^+ \nonu
\A\A - \hat g^{\mu\nu} \partial_\mu \chi^-
\partial_\nu \chi^- + Q \hat R \chi^- + \sum_{j=1}^N
\hat g^{\mu\nu} \partial_\mu f^j \partial_\nu f^j \biggr).
\label{dilkinzero}
\ea
Actions with different values of $Q$ are related by an
${\rm O}(1, 1)$ transformation in $\chi^+, \chi^-$ field space and
are equivalent. Finally the parameter $\kappa$ is determined by
requiring the conformal invariance (independence on
$\hat g_{\mu\nu}$) of the quantum theory (\ref{dilkin}) or
(\ref{dilkinzero}) \cite{BICA}
\be
\kappa={N-24 \over 12}.
\label{kappa}
\ee
\par
Let us note that the translation invariant path integral measure
for these free fields defines the quantization of the dilaton
gravity. In terms of the original variables, the path integral
measure may be quite nonlinear, and the path integral region
(values of the fields) may sometimes be peculiar.
Our attitude is that the dilaton gravity is nonlinear in the
original variable and is difficult to quantize.
Therefore we can use the free field representation
of the kinetic term (\ref{dilkin}) or (\ref{dilkinzero}) together
with the translation invariant measure as the definition of (the
kinetic term of) the quantum theory of the dilaton gravity.
\par
To discuss the physical states and the gauge invariant
operators of the theory, we compute the cohomology of the BRST
charge $Q_{\rm B}$ for the vanishing cosmological constant
$\mu = 0$. For the case of $0 \le N < 24$, the analysis of
refs.\ \cite{LIAN}, \cite{BILAL} applies with $c=N+1$.
Since we have not found the results for $N \ge 24$ in
the existing literature, we compute the BRST cohomology for
$N \ge 24$ as a direct extension of other cases
\cite{LIAN}, \cite{BILAL}, and find a result
similar to the case $0<N<24$ \cite{BILAL}.
In the case of $N\neq 24$
it is convenient to introduce the following notation for momentum
\be
p_I \equiv (-i\beta_{\chi}, -i\beta_{\Omega}, p_j).
\label{momentum}
\ee
corresponding to $X^I=(\hat \chi,\ \hat \Omega,\ f^j)$.
To raise or lower the indices we should use a metric in field space
$\eta_{IJ}=(-\kappa/|\kappa|, \kappa/|\kappa|, +1, \cdots, +1)$.
For $N=24$, we use $p_I \equiv (-i\beta_{+}, -i\beta_{-}, p_j)$ and
$\eta_{IJ}=(+1, -1, +1, \cdots, +1)$.
\par
Let us first describe the chiral (open string) cohomology
states for $N > 24$. The absolute cohomology consists
of $Q_{\rm B}$ invariant states modulo
$Q_{\rm B} \ket{\Lambda}$, where $\ket{\Lambda}$ is an arbitrary
state. The relative cohomology consists of $Q_{\rm B}$ invariant
states annihilated by the antighost zero mode $b_0$ modulo
$Q_{\rm B} \ket{\Lambda}$ with $\ket{\Lambda}$ annihilated by
$b_0$.
Usually the absolute cohomology states can be obtained
from the relative ones, so we will first obtain the relative
cohomology. For $(p_{\Omega}, p_j)\not=0$, the relative cohomology
consists of only the usual string excitations with vanishing
ghost number. Therefore the multiplicity of the states
with the ghost number $n$ is given by
\be
{\rm dim} H^n_{rel} = \left\{
\begin{array}{cl}
\delta_{n,0}P_N(R) & \qquad \mbox{for non-negative integer $R$,} \\
0 & \qquad \mbox{otherwise,}
\end{array} \right.
\label{stringexcit}
\ee
where the generating function of $P_N(R)$ is the usual
partition function for the oscillator excited states at the level $R$
and momentum $p_I$ has to satisfy the on-shell condition\footnote{
If we apply a naive hermiticity argument, we find that the momentum
$p_{\chi}=-i\beta_{\chi}$ for the field $\hat\chi$ should be purely
imaginary.}
\be
\prod_{m=1}^{\infty}(1-q^m)^{-N} =
\sum_{R=0}^{\infty}q^R P_N(R), \quad
{1 \over 2} p^I p^J \eta_{IJ} - i \sqrt{\kappa} p_{\chi} + R = 1.
\label{partitions}
\ee
Even if $(p_{\Omega}, p_j)=0$, the result is the same as
eq.\ (\ref{stringexcit}) unless $p_\chi = 0$ or $-2i\sqrt{\kappa}$.
Nontrivial cohomology similar to the discrete states in the $c=1$
gravity appear only for the following few cases:
\begin{enumerate}
\item For the $p_I = 0$,
we have $N+1$ states with vanishing ghost number
and one state with ghost number $-1$
\be
\alpha_{-1}^{\Omega} \ket{p_I=0}, \quad
\alpha_{-1}^{j} \ket{p_I=0}, \quad b_{-1} \ket{p_I=0},
\label{zeromom}
\ee
where $\ket{p_I}$ is the state with momentum $p_I$ annihilated by
all positive mode oscillators as well as $b_0$.
\item For $p_I\equiv(p_{\chi},p_{\Omega},p_j)
= (-2i\sqrt{\kappa}, 0, 0, \cdots, 0)$,
we have $N+1$ states with vanishing ghost number
and one state with ghost number $+1$
\be
\alpha_{-1}^{\Omega} \ket{p_I}, \quad \alpha_{-1}^{j} \ket{p_I},
\quad c_{-1} \ket{p_I}.
\label{twomom}
\ee
\end{enumerate}
Next we consider the relative cohomology for $N=24$.
In this case the result for the generic values of $p_I$ is the same
as in eqs.\ (\ref{stringexcit}) and (\ref{partitions})
except for the different on-shell condition
$-{1 \over 2} \beta_+(\beta_++Q)+{1 \over 2} \beta_-(\beta_--Q)+
{1 \over 2} p_j^2 + R = 1$.
The nontrivial cohomology classes exist only for $p_I=0$ or
$p_+=-p_-=i Q, p_j=0$
\ba
 (\alpha_{-1}^+ - \alpha_{-1}^-) \ket{p_I=0},
& \alpha_{-1}^{j} \ket{p_I=0},  \quad b_{-1} \ket{p_I=0}, \nonu
(\alpha_{-1}^+ + \alpha_{-1}^-) \ket{p_+=i Q},
& \alpha_{-1}^{j} \ket{p_+=i Q}, \quad
c_{-1} \ket{p_+=i Q}.
\label{kzerocohomo}
\ea
\par
In the present case it can be shown that all states in the
relative cohomology are non-trivial states of the absolute
cohomology. The remaining states in the absolute
cohomology are obtained by multiplying them by the ghost zero mode
$c_0$ and by adding certain terms if necessary \cite{LIAN}.
Namely, we find
\be
H^n_{abs} \simeq H^n_{rel} \oplus c_0 H^{n-1}_{rel}.
\label{abscoh}
\ee
\par
We also obtain the closed string cohomology
in the same way as in ref.\ \cite{WIZW}.
The relative cohomology consists of $Q_{\rm B} + \bar Q_{\rm B}$
invariant states modulo $(Q_{\rm B} + \bar Q_{\rm B})
\ket{\Lambda}$, where $\ket{\Lambda}$ is an arbitrary state.
The relative cohomology consists of $Q_{\rm B} + \bar Q_{\rm B}$
invariant states annihilated by the antighost zero modes
$b_0,\ \bar b_0$ modulo $(Q_{\rm B} + \bar Q_{\rm B})
\ket{\Lambda}$ with $\ket{\Lambda}$ annihilated by
$b_0,\ \bar b_0$. States in the closed string absolute (relative)
cohomology ${\cal H}^{n}_{\rm abs}$ (${\cal H}^{n}_{\rm rel}$) are
obtained by taking tensor products of two chiral absolute
(relative) cohomology states as left movers and right movers.
An important concept in the closed string cohomology
is the semi-relative cohomology, which consists of
$Q_{\rm B} + \bar Q_{\rm B}$ invariant states annihilated by
$b_0 - \bar b_0$ modulo $(Q_{\rm B} + \bar Q_{\rm B})
\ket{\Lambda}$ with $\ket{\Lambda}$ annihilated by
$b_0 - \bar b_0$ \cite{WIZW}.
In the present case it can be shown that
\be
{\cal H}^n_{semi} \simeq {\cal H}^n_{rel}
\oplus (c_0 + \bar c_0) {\cal H}^{n-1}_{rel}.
\label{closedabscoh}
\ee
\par
%
%%%%%%%  Cosmological Term  %%%%%%%%%%%%%%%%%%%%%%%%%%%%%%%%%%%%%%%%
%
The cosmological term is chosen from these BRST cohomology
classes by imposing the additional requirement \cite{BICA} that it
should be of the same form as the original cosmological term in the
limit of weak gravitational coupling ${\rm e}^{\phi}\rightarrow 0$.
We obtain uniquely\footnote{We have fixed the integration constant
in the field transformation (\ref{fieldredef}) between $\omega$ and
$\Omega$ by the asymptotic form
$\Omega \rightarrow -(\omega^2+(a-\kappa)\ln\omega)/(2\kappa)+
O(1/\omega^2)$.}
\be
S_{cosm}=\left\{
\begin{array}{ll}
{\mu \over \pi} \int d^2 z \sqrt{\hat g}
\e^{{1 \over \sqrt{|\kappa|}}(-\hat \chi+\hat \Omega)}
& \qquad \mbox{for $\kappa \neq 0$,} \\
{\mu \over \pi} \int d^2 z \sqrt{\hat g}
\e^{-{1 \over Q}(\chi^+ + \chi^-)}
& \qquad \mbox{for $\kappa = 0$.}
\end{array} \right.
\label{cosmdil}
\ee
The particularly interesting is the case of no additional matter
fields, i.e.,\ $N=0$ ($\kappa=-2$).
By identifying $\kappa=-2$ in eq.\ (\ref{dilkin}),
we find that the dilaton gravity system without additional matter
corresponds to the case of the Liouville gravity coupled to
the $c=1$ conformal matter. The only distinction is that the $c=1$
matter comes from the dilaton degree of freedom, and that the
resulting free boson $\hat \Omega$ has negative metric.
To compare with the physical operators in the Liouville gravity,
we should rotate the free boson to purely imaginary values
$\hat \Omega=i\bar \Omega$ and identify it with the usual free
boson with the positive metric. Then we find that the two-momentum
of this cosmological term is precisely the simplest discrete value
\cite{WITTEN} $(\beta_{\chi},p_{\bar\Omega})
=(-ip_{\chi},ip_{\Omega})=(-1/\sqrt{2}, 1/\sqrt{2})$.
\par
%
%%%%%%%  Correlation Functions  %%%%%%%%%%%%%%%%%%%%%%%%%%%%%%%%%%%%
%
Among these BRST cohomology classes, we shall take
momentum eigenstates
(gravitationally dressed tachyon vertex operators in the
language of the Liouville gravity), and
compute their correlation functions.
In the case of $N \neq 24$
\ba
O_p = \int d^2z \sqrt{\hat g} \, \e^{\beta_{\chi} \hat \chi
+ \beta_{\Omega}\hat \Omega+ip_j f^j}, \A\A \nonu
-{1 \over 2} \beta_{\chi}(\beta^{\chi}+2\sqrt{|\kappa|})
-{1 \over 2} \beta_{\Omega}\beta^{\Omega}+{1 \over 2}p_j p^j \A = \A 1.
\label{gaugeinvop}
\ea
The $(N+2)$-momentum of the cosmological term is
$q_I= ( i/\sqrt{|\kappa|}, -i/\sqrt{|\kappa|},0, \cdots, 0)$ and
that of the background source is
$Q^I=(- 2 i \sqrt{|\kappa|}, 0, 0, \cdots, 0 )$.
\par
It has been noted that the integration over the zero mode of the
Liouville field has to be done without using perturbation theory
in terms of the cosmological constant \cite{GOULI}, \cite{DFKU}.
By the same token, we perform the integration over the zero mode
of the transformed Liouville field $\hat \chi$ without expanding
in powers of the cosmological constant in the action.
After integrating over the zero mode of $\hat \chi$ along the real
axis,  we integrate over the zero mode of $\hat \Omega$
along the imaginary axis to obtain the momentum conservation delta
function\footnote{An arbitrary linear combination of two variables
$\hat \chi$ and $\hat \Omega$ can be used instead of
$\hat \chi$ and its orthogonal combination instead of
$\hat \Omega$ to give the same result, provided the
latter variable is integrated along the imaginary axis.}. In this
way we find the path integral with cosmological term insertions
\ba
\VEV{\prod_{k=1}^n {O_{p_k}}}
\A = \A \int {[d\tau] \D_{\hat g} \hat\chi \D_{\hat g} \hat\Omega
     \D_{\hat g} f \over V_{\rm CKV}} \,
     \e^{- S[ \hat\chi, \hat\Omega, f, \hat g]} \,
     O_{p_1} \cdots O_{p_n} \nonu
\A = \A (2\pi)^{N+1} \delta^N \left( \sum_{k=1}^n p_k \right)
     \delta \left( \sum_{k=1}^n \beta_{\Omega k}
     + {s \over \sqrt{|\kappa|}} \right)
     \sqrt{|\kappa|} \, \Gamma(-s)
     \tilde A (p_1, \cdots, p_n), \nonu
\tilde A = \int \A\A\!\!\!\!\!\!\! {[d\tau] \prod_{k=1}^n
\left[ d^2 z_k \sqrt{\hat g(z_k)} \right] \over V_{\rm CKV}}
\VEV{\prod_{k=1}^n \e^{ip_k \cdot {\tilde X}(z_k)}
\left( {\mu \over \pi} \int d^2 w \, \sqrt{\hat g} \,
\e^{i q \cdot \tilde X} \right)^s}_{\tilde X},
\label{corxphi}
\ea
where $s$ is the number of cosmological term insertions
and the $(N+2)$-momentum conservation reads
\be
s =  \sqrt{|\kappa|}  \sum_{k=1}^n \beta_{\chi k}
- 2 \kappa(1-h), \qquad
\sum_{k=1} p_{k}^I + s q^I = -Q^I(1-h).
\label{defofs}
\ee
The expectation value in eq. (\ref{corxphi}) denotes the functional
integral over the nonzero modes $\tilde X^I$.
As is usual in the Liouville gravity, the zero mode integration is
done assuming $s<0$.
If we continue analytically (in external momenta and/or the central
charge) to a nonnegative integer $s$, we can integrate
the non-zero mode functional integral in $\tilde A$ by means of the
usual free field contractions.
Finally we analytically continue the result to desired values of $s$.
\par
For the case of $N \not= 24, 0$, we find the
three-point correlation functions on the sphere
\be
\tilde A (p_1, p_2, p_3)
= \Bigl[ \mu \Delta(1 + p_1 \cdot q) \Delta(1 + p_2 \cdot q)
\Delta(1 + p_3 \cdot q) \Bigr]^s,
\label{threepoint}
\ee
where $\Delta(x) \equiv \Gamma(x) / \Gamma(1-x)$.
We see that the three-point functions exhibit singularity
structures corresponding to the physical states obtained in the
BRST analysis. However, the peculiarity of the correlation
functions is that these singularities are all raised to a power
of the cosmological term insertions $s$.
It is not difficult to see the origin of these singularities from
the operator product expansion. The cosmological term has the
null momentum $q_I$ and the integration
over the positions of the cosmological term operators is
factorized. The singularities are due to the short distance
singularities between these cosmological term operators and the
inserted gauge invariant operators.
We can obtain two-point function from the three-point
function if we put one of the momenta to that of the
cosmological term $q_I$, replace $s$ by $s-1$, multiply
by $-1/\pi$, and integrate in $\mu$ once.
Similarly, the one-point function is obtained by repeating the same
procedure for the two-point function.
We find that these correlation functions vanish for
$s\in {\bf Z}, s>0$ because $1 + q\cdot q = 1$. Therefore the
analytic continuation gives vanishing two- and one-point functions.
\par
For the $N=24$ case we obtain the three-point correlation functions
on the sphere
\be
\VEV{\prod_{k=1}^3 {O_{p_k}}}
= (2\pi)^{25} \delta^{24} \left( \sum_{k=1}^3 p_k \right)
  \delta \left( \frac{1}{Q} \sum_{k=1}^3 \left(\beta_{+ k}
  - \beta_{- k} \right) + 2 \right)\, \Gamma(-s)
  \tilde A (p_1, p_2, p_3),
\label{threepointN24}
\ee
with the same amplitude $\tilde A$ in eq.\ (\ref{threepoint})
and the number of cosmological term insertion $s$ is given by
\be
s =  \frac{Q}{2} \sum_{k=1}^3 \left(\beta_{+ k}
 + \beta_{- k} \right).
\label{defofsn}
\ee
For $N=24$ the momenta of the cosmological term and
the background source are
$q_I = \left( {i \over Q}, {i \over Q}, 0, \cdots, 0 \right)$
and $Q_I=(- iQ, iQ, 0, \cdots, 0 )$ respectively,
and momentum conservation holds in the same form as
eq.\ (\ref{defofs}). Since $q \cdot q = 0$ is still valid,
one and two point functions vanish in this case too.
\par
In the case of $N=0$, we can obtain correlation functions for
arbitrary numbers $n$ of gauge invariant operators
(\ref{gaugeinvop}) on the sphere.
Since the cosmological term in this case is at the
special discrete momentum, the correlation functions become
singular due to their insertions \cite{DFKU}.
Therefore we need to keep the momentum $q_I$ of the
cosmological term at a generic value to regularize the
correlation functions.
After an analytic continuation in $s$ analogous to the
$c=1$ Liouville gravity theory, we obtain the $n$-point
correlation functions. In the case of $N=0$, it is important
to distinguish two solutions of the on-shell condition
(\ref{gaugeinvop})
\be
\beta_{\chi}^{(\pm)}=-\sqrt{2}\pm\beta_{\Omega}.
\label{chirality}
\ee
Following the terminology of Liouville gravity \cite{DFKU},
we call the solution with the $+$ $(-)$ sign as positive (negative)
chirality.
We find the correlation functions of $n$ gauge invariant
operators with chiralities $(-,+, \cdots, +)$
\ba
\VEV{\prod_{k=1}^n O_{p_k}} \A=\A
2\pi \delta \left( {1 \over \sqrt2}\sum_{k=1}^n
(\beta_{\chi k} + \beta_{\Omega k}) + 2
\right) \Gamma(-s) \nonu
\A\A \times \left[\mu\Delta (-\rho) \right]^s
{\pi^{n-3}  \over \Gamma(n+s-2)}
\prod_{k=2}^n \Delta(1 - \sqrt{2} \beta_{\Omega k}),
\label{dilcor}
\ea
where the regularization parameter $\rho = q \cdot q / 2$ should
be set to zero eventually.
The correlation functions are very similar to the Liouville gravity
case, but are singular due to the discrete momentum for the
cosmological term. We can absorb the divergence as a
renormalization of the cosmological constant
\be
\mu_r=\mu \Delta(-\rho).
\label{rencosm}
\ee
Correlation functions vanish identically for chirality
configurations other than the above one. When computing these
correlation functions, one should keep the regularization
parameter $\rho$ to be nonvanishing in order to avoid ambiguous
results.
\par
We hope that the above results on the physical states and
the correlation functions
may serve as a first step for a full quantum
treatment of the dilaton gravity beyond the usual semi-classical
approximations. Let us emphasize that the continuum approach is
flexible enough to study the dilaton gravity, whereas the powerful
method of matrix models is yet to be applied to this case.
However, when we are writing this paper, we received an interesting
preprint which computed the partition function with the tachyon
background for the $c=1$ quantum gravity by using matrix model
approach \cite{DIMOPL}. We are currently studying to
apply their results to our case of $N=0$ dilaton gravity.
The wave functions of the universe and possible consequences on the
black hole evaporation are being studied.
\par
\vspace{5mm}
We thank T. Mishima, A. Hosoya, and A. Nakamichi for a useful
discussion and for informing us their results prior to publication.
One of us (N.S.) thanks H. Kawai and C. G. Callan for a discussion.
This work is supported in part by Grant-in-Aide for Scientific
Research for Priority Areas from the Ministry of Education,
Science and Culture (No. 04245211).
\vspace{5mm}

\begin{thebibliography}{100}
%
\bibitem{CGHS} C.G. Callan, S.B. Giddings, J. Harvey and
        A. Strominger, {\it Phys.\ Rev.\ }{\bf D45} (1992) R1005.
\bibitem{RUTS} J. Russo and A.A. Tseytlin, Stanford preprint
        SU-ITP-92-2 (1992);
        J. Russo, L. Susskind and
        L. Thorlacius, Stanford preprint SU-ITP-92-4 (1992);
        T. Banks, A. Dabholkar, M. Douglas, and M. O'Loughlin,
        {\it Phys.\ Rev.\ }{\bf D45} (1992) 3607;
        S.W. Hawking, {\it Phys.\ Rev.\ Lett.\ }{\bf 69} (1992) 406;
        D. Cangemi and R. Jackiw,
        {\it Phys.\ Rev.\ Lett.\ }{\bf 69} (1992) 233;
        T.T. Burwick and A.H. Chamseddine, Z\"urich preprint
        ZU--TH--4/92 (1992);
        S.W. Hawking and J.M. Stewart, DAMTP preprint (1992).
\bibitem{STR} A. Strominger, Santa Barbara preprint UCSBTH--92--18
        (1992).
\bibitem{BICA} A. Bilal and C. Callan,
        Princeton preprint PUPT--1320 (1992);
        S.P. de Alwis, Colorado preprints
        COLO-HEP-280, 284, 288 (1992);
        S.B. Giddings and A. Strominger, Santa Barbara preprint
        UCSB-TH-92-28 (1992);
        A. Mikovi\'c, Queen Mary preprint QMW/PH/92/12 (1992).
\bibitem{BILAL} A. Bilal,
        {\it Phys.\ Lett.\ }{\bf B282} (1992) 309.
\bibitem{WITTEN} E. Witten, {\it Nucl.\ Phys.\ }{\bf B373}
        (1992) 187; J. Avan and A. Jevicki,
        {\it Phys.\ Lett.\ }{\bf B266} (1991) 35;
        D. Minic, J. Polochinski and Z. Yang,
        {\it Nucl.\ Phys.\ }{\bf B369} (1992) 324;
        S. Das, A. Dhar, G. Mandal and S. Wadia,
        {\it Mod.\ Phys.\ Lett.\ }{\bf A7} (1992) 71, 937;
        I.R. Klebanov and A.M. Polyakov,
        {\it Mod.\ Phys.\ Lett.\ }{\bf A6} (1991) 3273;
        Y. Matsumura, N. Sakai and Y. Tanii,
        Tokyo Inst.\ of Tech.\ and Saitama preprint
        TIT/HEP--186, STUPP--92--124 (1992).
\bibitem{GOULI} M. Goulian and M. Li,
        {\it Phys.\ Rev.\ Lett.\ }{\bf 66} (1991) 2051;
        A. Gupta, S. Trivedi and M. Wise,
        {\it Nucl. Phys.\ }{\bf B340} (1990) 475.
\bibitem{DFKU} P. Di Francesco and D. Kutasov,
        {\it Phys.\ Lett.\ }{\bf B261} (1991) 385;
        {\it Nucl. Phys.\ }{\bf B375} (1992) 119;
        Y. Kitazawa, {\it Phys.\ Lett.\ }{\bf B265} (1991) 262;
        N. Sakai and Y. Tanii,
        {\it Prog.\ Theor.\ Phys.\ }{\bf 86} (1991) 547;
        V.S. Dotsenko, {\it Mod.\ Phys.\ Lett.\ }{\bf A6}
        (1992) 3601.
\bibitem{POLYAKOV} A.M. Polyakov, {\it Phys.\ Lett.\ }{\bf B103}
        (1981) 207; D. Friedan, in {\it Recent Advances in Field
        Theory and Statistical Mechanics}, eds.\ J.B. Zuber and
        R. Stora, (North-Holland, Amsterdam, 1984).
\bibitem{DDK} F. David, {\it Mod.\ Phys.\ Lett.\ }{\bf A3} (1988)
        1651; J. Distler and H. Kawai,
        {\it Nucl.\ Phys.\ }{\bf B321} (1989) 509.
\bibitem{LIAN}B.H. Lian and G.J. Zuckerman,
        {\it Phys.\ Lett.\ }{\bf B254} (1991) 417;
        {\bf B266} (1991) 21;
        P. Bouwknegt, J. McCarthy and K.Pilch,
        {\it Commun.\ Math.\ Phys.\ }{\bf 145} (1992) 541;
        N. Ohta, Osaka preprint OS--GE 26--92 (1992).
\bibitem{WIZW} E. Witten and B. Zwiebach,
        {\it Nucl.\ Phys.\ }{\bf B377} (1992) 55.
\bibitem{DIMOPL} R. Dijkgraaf, G. Moore and R. Plesser,
        Princeton preprint IASSNS--HEP--92/48 (1992).
\end{thebibliography}
\end{document}